\documentclass[aps,twocolumn,superscriptaddress,prl]{revtex4}
\usepackage{amsmath,latexsym,epsf}
\usepackage{graphicx}
\usepackage{subfigure}

\begin{document}

\title{Critical Dynamics of Superconductors in the Charged Regime}

\author{Courtney Lannert}

\affiliation{Department of Physics, Wellesley College, 106 Central St., Wellesley, MA 02481-8203}

\author{Smitha Vishveshwara}

\affiliation{Department of Physics, University of Illinois at Urbana-Champaign, 1110 W. Green St., Urbana, IL 61801-3080}

\author{Matthew P.A. Fisher}

\affiliation{Kavli Institute for Theoretical Physics, University of California, Santa Barbara, CA 93106-4030}

\date{\today}

\begin{abstract}

We investigate the finite temperature critical dynamics of 
three-dimensional superconductors in the charged regime, described
by a transverse gauge field coupling to the superconducting order
parameter. Assuming relaxational dynamics for both the 
order parameter and the gauge fields, within a dynamical 
renormalization group scheme, we find a new dynamic universality
class characterized by a finite fixed point ratio between the 
transport co-efficients associated with the order parameter and 
gauge fields, respectively. We find signatures of this 
universality class in various measurable physical quantities, and in 
the existence of a universal amplitude ratio formed by a combination
of physical quantities.  
\end{abstract}

\vspace{0.15cm}

\maketitle

Close to the critical temperature $T_c$ of the normal-superconductor 
transition, in a regime determined by the Ginzburg criterion \cite{gc,FFH}, 
 order parameter fluctuations dictate  critical properties.
For decades, the effect of the charge of the superconducting order 
parameter in this regime in three dimensions  has formed the topic of keen
 study. While for strongly type-I materials, 
the coupling of the order parameter to transverse gauge field 
fluctuations is expected to render the transition first order \cite{HLM74}, 
it is well-established that strongly type-II materials should exhibit a 
\emph{continuous} phase transition, and that sufficiently close to $T_c$,
the charge of the order parameter field is relevant \cite{DH81}. While the 
exact location of the boundary between these two types of behavior is 
still the subject of investigation \cite{MHS02}, the static critical
 properties of the charged-XY universality class are reasonably
 well-understood.
With the discovery of superconducting compounds with large 
critical temperatures and short coherence lengths the critical
 regime of this transition is now potentially accessible to 
experimental investigation.

In this Letter, we investigate the less-addressed issue \cite{AG01,LW98}
of the dynamics of the three-dimensional normal-superconductor transition in 
the charged regime. In the well-studied case of superfluid $\mbox{He}^4$, 
the coupling of the order parameter to a conserved energy density field 
has non-trivial effects on critical dynamics \cite{HH77}.
 Analogously, we propose that the
 coupling of the superconducting order parameter to relaxational transverse
 gauge field fluctuations leads to qualitatively new dynamics 
 characterized by a universal
ratio ${\cal C}$ between the zero wave-vector part of the characteristic frequencies for the
dynamics of  the order parameter $\psi_{\alpha}$ 
and that of the gauge field $\mathbf{A}$ at the critical point:
\begin{equation}
{\cal C} \equiv \lim_{k\rightarrow 0} \frac{\omega_{\psi} (k)}{\omega_{A} (k)} = \rm{const}.
\label{uar}
\end{equation}
Thus we propose the possibility that
the strong coupling of these two fields causes them to relax 
in the same fashion at the critical point,
 with a single new dynamic exponent $z$.
In what follows, employing dynamic renormalization group (RG)
techniques, we show these features to hold within the context of
a particular model. We discuss the universal properties obtained from the
 model and the behavior of measurable physical quantities.

\emph{\bf The Model- } 
As a starting point for modeling dynamics, we employ the 
standard Ginzburg-Landau free energy used to define
the finite temperature static critical properties of a three-dimensional
superconductor coupled to a transverse electromagnetic field \cite{Tinkham}, generalized to 
$N$ complex species of matter field, given by
\begin{eqnarray}
F &=& \int d^d x \left[\sum_{\alpha} \left| \left( \nabla - i\sqrt{g} \mathbf{A} \right) \psi_\alpha 
\right|^2 + r \sum_{\alpha} | \psi_\alpha |^2 + \right. 
\nonumber \\
& & \left. + u \left(\sum_{\alpha} |\psi_\alpha|^2\right)^2 + \frac{1}{2} |\nabla \times \mathbf{A} |^2 \right],
\label{gl}
\end{eqnarray}
where $\psi_\alpha$ (with $\alpha \in {1, 2, ... N}$) is the generalized order parameter and $\mathbf{A}$ is a fluctuating
 massless gauge field. The effective charge is given by
 $\sqrt{g} = e^* \sqrt{4\pi}/\hbar c $ and deviations from criticality
are measured by $r = 2m^*(T-T_c)/\hbar^2T_c$ \footnote{Where $e^*$ is 
the charge of the superconducting field, for instance $e^*=2e$ in 
standard BCS theory.}. For $N=1$ and $d=3$, this model describes the low-energy excitations of a
bulk charged superfluid in the regime where charge-density fluctuations are gapped
 at a high-energy (the plasmon gap), but remain coupled to transverse
 electromagnetic fluctuations. 

As originally obtained in Ref.\cite{HLM74}, a one-loop static RG analysis of Eq.(\ref{gl}) in $d=4-\epsilon$ dimensions shows that no non-trivial charged fixed point exists for $2N<n_c=365.9$. However, for a range of parameters, more sophisticated
methods indeed find a continuous phase transition for $N=1$ \cite{DH81,K82-9,HT96},
describing, presumably, the second order transition found in many
materials. Many of the salient features of this ``charged-XY" universality
class are captured by Eq.(\ref{gl}) with $N>N_c \simeq 183$, for instance the
fact that the anomalous dimension of the order parameter field is negative. 
Given that the $N$-component model does provide insight on the statics
of the charged superconducting transition, here we make the reasonable 
assumption that when the free-energy defined by  Eq.(\ref{gl}) is
augmented by appropriate equations of motion, the model captures the
basic features of the critical dynamics of the transition as well.

For completeness, we note that the free energy density is invariant under the following transformations:
\begin{eqnarray}
&&\psi_{\alpha}^a \rightarrow \delta_{ab} {\cal O}_{\alpha \beta} \psi_{\beta}^b, \\
&&\left\{ \begin{array}{lcl}
\psi_{\alpha}^a &\rightarrow& e^{i(\sqrt{g}\Lambda(\vec{r}) + \theta)\sigma_{ab}^y} \delta_{\alpha \beta} \psi_{\beta}^b \\
\mathbf{A}(\mathbf{r}) &\rightarrow& \mathbf{A}(\mathbf{r}) + \nabla \Lambda(\mathbf{r})
\end{array} \right., \label{gauge_symmetry}
\end{eqnarray}
where $\alpha,\beta \in \{1,..., N\}$, $a,b \in \{1,2\}$, ${\cal O}$ is an $N\times N$ orthogonal matrix, $\theta \in [0,2\pi]$, and $\sigma_{ab}^y$ is the usual $2\times2$ Pauli matrix. $\psi_{\alpha}^1$ and $\psi_{\alpha}^2$ are the real and imaginary parts, respectively, of the $\alpha$-th complex field.
The term in Eq.(\ref{gl}) with coupling constant $u$ is the only quartic matter-matter interaction allowed by these symmetries. 

The simplest equations of motion
augmenting the statics described by  Eq.(\ref{gl})
 are relaxational in both $\psi_\alpha$
and $\mathbf{A}$. In the presence of external fields $h_{\psi \alpha}$ and
$h_{\mathbf{A}}$ coupling to $\psi_\alpha$ and $\mathbf{A}$
respectively, they take the form
\begin{eqnarray}
\partial_t \psi_{\alpha} &=& -\Gamma_{\psi} 
\left(\frac{\delta F}{\delta \psi_{\alpha}^*} - h_{\psi \alpha}\right)
+ \eta_\alpha, \label{psi_t} \\
\partial_t A_i &=& -\Gamma_{A} \left(\frac{\delta F}{\delta A_i} -
h_{Ai}\right) + \zeta_i. \label{A_t}
\end{eqnarray}
Here, $\Gamma_{\psi}$ and $\Gamma_A$ are transport co-efficients associated
with $\psi_{\alpha}$ and $\mathbf{A}$ respectively.
The fields $\eta_{\alpha}, \eta^*_{\alpha}$ and {\boldmath$\zeta$} are white noise correlated,
and ensure that the fluctuation-dissipation theorem is
satisfied. Thus, they obey the constraints
\begin{eqnarray}
\langle \eta_{\alpha} (\mathbf{r},t) \eta_{\beta}^*(\mathbf{r}',t')\rangle 
&=& 2T\Gamma_{\psi}\delta_{\alpha \beta}\delta^3(\mathbf{r}-\mathbf{r}')\delta(t-t'), 
\label{eta} \\
\langle \eta_\alpha(\mathbf{r},t) \rangle &=& \langle \eta_\alpha(\mathbf{r},t) 
\eta_\beta(\mathbf{r}',t') \rangle = 0, \\
\langle \zeta_i (\mathbf{r},t) \zeta_j (\mathbf{r}',t')\rangle 
&=& 2T\Gamma_{A}\delta_{ij}\delta^3(\mathbf{r}-\mathbf{r}')\delta(t-t'), 
\label{zeta} \\
\langle \zeta_i (\mathbf{r},t) \rangle &=& 0.
\end{eqnarray}
In fact, if we assume that the normal state of the system is a metal,
Eq.(\ref{A_t}) can be derived from the low-frequency form of Maxwell's equation
\begin{equation}
\mathbf{\nabla}\times\mathbf{B}=4\pi\mathbf{J}/c+\partial_t\mathbf{E}/c,
\label{Maxwell}
\end{equation}
in the gauge $A_0=0$, where the electric and magnetic fields are given 
by $\mathbf{E}=-\partial_t\mathbf{A}/c$ and 
$\mathbf{B}=\mathbf{\nabla}\times\mathbf{A}$, respectively.
The net current $\mathbf{J} = \mathbf{j}_s + \mathbf{j}_n$ has a superfluid
component $\mathbf{j}_s= -\frac{\delta}{\delta \mathbf{A}} |(\nabla - i\sqrt{g}\mathbf{A})\psi_{\alpha}|^2$ and a normal component $\mathbf{j}_n$.
The average normal current is given by $\sigma_n \mathbf{E}$, where
$\sigma_n$ is the normal conductivity. Thermal fluctuations of the normal
fluid give rise to the noise in Eq.(\ref{A_t}). With these 
assumptions, one can retrieve Eq.(\ref{A_t}) from Eq.(\ref{Maxwell}) in the limit
$\omega\rightarrow 0$, and identify the inverse transport co-efficient,
$\Gamma_A^{-1}$, with the bare conductivity, $\sigma_n$.

A complete model for the dynamics requires identification of
all conserved quantities and Poisson-bracket relations applicable
to the normal-superconductor system in the charged regime \cite{HH77}.
 Even in the uncharged
regime, one might expect non-dissipative coupling
of the order parameter to a combination of energy and mass density
to exhibit model E dynamics as in superfluid helium \cite{HH77}. 
However, hydrodynamic analyses show that in the presence of impurities 
(which is implicit in the assumption of finite conductivity well
within the normal state) this coupling does not survive \cite{MFSV}, as 
indicated by the absence of second sound modes in actual superconducting
systems. However, in principle, a conserved energy density mode could 
couple to the order parameter mode via non-linear interactions,
leading to model C dynamics in the uncharged superconductor \cite{MFSV}.
Likewise, in the charged regime, we do not expect any 
non-dissipative coupling of energy-mass density to the order parameter, 
due to the presence of impurities in real samples of interest. 
However, the possibility
of non-linear interactions with conserved quantities cannot 
be completely eliminated. The model we employ consisting of
the order parameter and gauge field modes alone, each with relaxational
dynamics, is the simplest, but hitherto unexplored, possibility.
   
\emph{\bf RG analysis- } The equations of motion Eqns.(\ref{psi_t},\ref{A_t}) allow a 
dynamical RG analysis which we now detail (also see, for example, Ref.\cite{HH77}).
 The effective charge $\sqrt{g}$ and the coupling constant $u$ of
 Eq.(\ref{gl}) are treated perturbatively, as is the deviation from four
 dimensions, $\epsilon = 4-d$ in order
to avoid infrared divergences \cite{RG}.
Because the characteristic electron speeds are small compared to the speed of
light, the low-energy theory of the system need not be relativistically
invariant. This leads to the residual gauge symmetry contained in
Eq.(\ref{gauge_symmetry}). (This has been pointed out previously in, for
instance, Ref.~\cite{FG92}.) To avoid divergences in functional integrals
resulting from the fact that multiple choices of gauge lead to the same magnetic
field configuration, we perform the Fadeev-Popov procedure and 
add a term $1/(2\lambda)(\nabla \cdot \mathbf{A})^2$ to the 
free energy (Eq.(\ref{gl}))\cite{ps}. Fixing the value of 
$\lambda$ corresponds to making a choice of gauge, and for the
remainder of the paper we work with $\lambda =1$ 
(analogous to the Feynman gauge in quantum electrodynamics). Regarding the
dynamics, since our focus is on the manner in which the relaxational
rates of the fields $\psi$ and \textbf{A} affect one another, 
we  rescale the theory and write the equations of motion in terms of the 
ratio of transport co-efficients $\Gamma \equiv  \Gamma_{\psi}/\Gamma_{A}$. 

We perform the standard RG procedure, integrating out modes in a momentum
shell $\Lambda/b < |\mathbf{k}|<\Lambda$, where $\Lambda$ is a high momentum
cut-off,
and all frequencies, followed by a rescaling of space and time: 
$\mathbf{r} \rightarrow b\mathbf{r}$ and $t \rightarrow b^z t$.  
To one-loop, and ${\cal O}(\epsilon)$, the $\beta$-functions for $u$ and $g$ are (as in Ref.\cite{HLM74}):
\begin{eqnarray}
\frac{du}{d\ln b} &=& \epsilon u -2(N+4)u^2 - 3g^2 + 6gu, \\
\frac{dg}{d\ln b} &=& \epsilon g -\frac{N}{3} g^2 = g(\epsilon - \eta_A),
\end{eqnarray}
yielding Gaussian ($u^*=g^*=0$), XY ($u^*\neq 0$, $g^*=0$), and charged-XY
($u^*\neq0$, $g^*\neq0$) fixed points for $N\geq 183$. At the charged-XY fixed
point of interest, $g^*=3\epsilon/N$ to leading order in $\epsilon$. 
We find that the anomalous dimensions of the order parameter and 
gauge field at this order (and in our choice of gauge) are:
\begin{equation}
\eta_{\psi} = -2g \;\; ; \;\; \eta_A = Ng/3,
\end{equation}
and note that $\eta_A=\epsilon=4-d$ at the critical point, as required by gauge invariance \cite{HT96}. 

We fix the
exponent $z$ by requiring that the time derivative term in the equation of
motion for \textbf{A} return to its ``bare" form. The same feat in the
equation of motion for $\psi$ is accomplished by allowing the ratio $\Gamma$
to flow. We find that
\begin{eqnarray}
1 &=& b^{2-\eta_A-z}\left( 1+ g\frac{N}{2\Gamma}\ln b \right) \label{z}, \\
\Gamma' &=& b^{z-2+\eta_{\psi}}\Gamma \left( 1 - g\frac{\Gamma}{\Gamma+1}\ln b\right), \label{Gamma_flow}
\end{eqnarray}
giving a one-loop $\beta$-function for $\Gamma$:
\begin{equation}
\frac{d\Gamma}{d\ln b} = g\Gamma\left(\frac{N}{2\Gamma} - \frac{\Gamma}{\Gamma+1}-2-\frac{N}{3} \right). \label{Gamma_beta}
\end{equation}

At the ${\cal O}(\epsilon)$ charged fixed point, Eq.(\ref{Gamma_beta}) has a stable fixed point
solution:
\begin{equation}
\Gamma^*=\frac{(N-12)+\sqrt{(N-12)^2+24N(N+9)}}{4(N+9)},
\end{equation}
which in general obeys $0\leq \Gamma^* < 1.5$ for all $N\geq 0$.
We note that $\Gamma =0$ ($\Gamma_\psi \rightarrow 0$ with $\Gamma_A$ fixed, for instance) is not a fixed point and that $\Gamma^{-1} =0$ ($\Gamma_A \rightarrow 0$ with $\Gamma_\psi$ fixed, for instance) is an unstable fixed point, at this order. Thus, we are led to conclude that, at least within a one-loop RG analysis, \emph{the critical dynamics of the charged 
superconductor is governed by a non-trivial fixed point}
wherein $\Gamma$ has a finite ratio, reflecting the fact that the dynamics 
of the order parameter and those of the gauge field are strongly coupled.
At the  charged-XY fixed point, we find that $z =2+\epsilon[3/(2\Gamma^*) -1]$
to leading order in $\epsilon$, implying that $z>2$.
While an accurate value of the dynamic exponent would require employing
more sophisticated treatments, our results certainly suggest
that near criticality, the system relaxes slower than expected for
diffusive dynamics.

\emph{\bf Physical Consequences- }
The most striking new feature of this fixed point (in contrast with the
 uncharged dynamics) is the existence of the universal fixed ratio, 
$\Gamma^*$. Physically, its existence requires that the order parameter
and gauge fields relax in the same fashion. In fact, using scaling arguments
\cite{HH77}, one can show that at the critical point, the ratio of the
characteristic frequencies at $k=0$ of the two fields is exactly the universal
amplitude ratio, identifying the constant ${\cal C}$ of Eq.(\ref{uar})
with $\Gamma^*$, \emph{i.e.},
\begin{equation}
{\cal C} \equiv \lim_{k\rightarrow 0}\frac{\omega_{\psi} (k)}{\omega_{A} (k)} = \Gamma^*.
\label{uar2}
\end{equation}
Here, the characteristic frequency of a field $Q$ is defined by
\begin{equation}
\omega^{-1}_Q(k)=i\chi_Q(k,\omega=0)
\left. \frac{\partial\chi_Q^{-1}(k,\omega)}{\partial\omega}\right|_{\omega=0},
\label{charfreq}
\end{equation}
where $\chi_Q$ is the dynamic linear response function, $\langle Q\rangle=\chi_Q h_Q$.
In our case, close to criticality, the characteristic frequency
has the scaling behavior $\omega_Q=\xi^{-z}\Omega_Q(k\xi)$
for both fields $\psi$ and $\mathbf{A}$, where $\xi$ is the 
divergent correlation length associated with the order parameter,
and $\Omega_Q$ is a universal function. In principle, the 
characteristic frequency for each field can be obtained by measuring
the static susceptibility and dynamic linear response function
associated with the field.

The dynamic linear response functions themselves carry valuable
information on critical dynamics. Close to $T_c$, they have 
the scaling form
\begin{equation}
\chi_Q(k,\omega)=\xi^{2-\eta_Q}f_Q(\omega\xi^z,k\xi),
\label{linresp}
\end{equation}
where, associated with each field $\psi_{\alpha}$ and $\mathbf{A}$,
$\eta_Q$ and $f_Q$ are the anomalous dimension and a 
scaling function, respectively.

Each of these functions is manifest in measurable
quantities. The order parameter response function, $\chi_{\psi}(k,\omega)$,
is the pair susceptibility appearing in  Josephson tunneling
experiments \cite{AGDJS70}. In principle, the
scaling behavior of $\chi_{\psi}$ in Eq.(\ref{linresp}) could be used to extract 
the dynamic exponent, $z$. Accessing $\chi_A(k=0,\omega)$ should be relatively straightforward, since it is related to the resistivity in linear response. The external field that couples to $\mathbf{A}$ is an applied current $\mathbf{j}_{ext}$, so that,  
$\langle\mathbf{A}\rangle=\chi_A\mathbf{j}_{ext}$, with
$\sigma(\mathbf{k}=0,\omega)= - i \chi^{-1}_A(\mathbf{k}=0,\omega)/\omega$. 
 Close to criticality and above $T_c$, we expect this to provide the dominant contribution to the
net conductivity and, using Eq.(\ref{linresp}), to diverge as 
$\sigma(k=0,\omega)=\xi^{z-2+\eta_A}{\cal G}(\omega\xi^z) \sim \xi^{z+2-d}$, where
${\cal G}$ is a scaling function. This is a consequence of the Josephson
scaling relation \cite{J69,HT96} which holds in both charged and uncharged
regimes, as a result of gauge invariance \cite{HT96}. 
Notably, in the charged regime, the exponent $z$ has a different 
value than in the uncharged case.
 
We see that various features of the dynamic universality
class of the charged superconductor appear in measurable quantities.
The definitive signature of this universality class would be in the 
extraction of the universal amplitude ratio, Eq.(\ref{uar2}).

\emph{\bf Concluding Remarks- }
The charged superconductor-normal
 transition is a continuing source of rich physics. 
Research on the static transition of the charged Ginzburg-Landau model indicates differing behavior 
for type-I and type-II superconductors as well as the relevance (sufficiently close to the continuous transition) of the charge of magnetic field fluctuations.
Numerical and analytical work in Ref.\cite{AG01} on the dynamics of this
 transition also seem to reveal distinctly different dynamic behavior between
strongly and weakly screened superconductors.
In Ref.\cite{LW98}, Monte Carlo studies of superconductor
dynamics in the vortex representation find that for strong magnetic screening, 
$z\approx 2.7$, in qualitative agreement with our result that
the order parameter dynamics in the charged regime is sub-diffusive.
 Here, as our key point, we suggest that for materials with continuous transitions from 
normal metal to superconductor, the dynamics in the charged regime
 will be governed by a new universality class.
Further analyses of all these issues are well in order.

In experiments, the charged regime of the superconductor-metal transition is not
easily accessible. The Ginzburg criterion indicates that materials with high critical
temperature, large anisotropy and extreme type-II behavior 
should manifest large regimes of fluctuations. However, within this fluctuation
regime, the region close to $T_c$ where the system crosses over
 to the regime of \emph{charged} fluctuations is often too narrow to access \cite{FFH}.
For instance, high $T_c$ materials such as $\mbox{YBa}_2 \mbox{Cu}_3 \mbox{O}_{7-d}$, while possessing large regimes of critical fluctuations,
are too strongly type-II to observe charged critical fluctuations.  However, 
weakly type-II materials with high $T_c$'s or granular texture, 
and moderate anisotropies could open up a window onto this new regime. We are
hopeful that an investigation of such materials will yield an understanding of
the effect of charge on the critical dynamics of this transition.

Many thanks to L. Balents and K. Wiese for valuable discussions, to
A. Ludwig, D. Scalapino, E. Fradkin, N. Goldenfeld and M. Randeria
for insightful comments, and to I. Herbut and Z. Te\v sanovi\' c for clarifying remarks on the
Josephson relation in the charged statics. This work was supported by the
grants NSF DMR-0210790, NSF PHY-9907949, NSF EIA01-21568, 
DOE DEFG02-96ER45434 and NSF PHY00-98353.

\end{document}